\documentclass[showpacs,aps,prd,nofootinbib,floatfix,amsmath,amssymb]{revtex4}
\usepackage{graphicx}
\usepackage{dcolumn}
\usepackage{bm}
\usepackage{subfigure}
\usepackage{color}
\usepackage{slashed}

\newcommand{\ra}{\rangle}
\newcommand{\la}{\langle}

\begin{document}


\title {$\tau^{-}\to \ell_i^{-} \ell_i^{+} \ell_j^{-} \bar{\nu}_{j} \nu_{\tau}$ decays with a magnetic dipole term}

\author{M. A. Arroyo-Ure\~na, E. D\'iaz, O. Meza-Aldama, G. Tavares-Velasco}

\affiliation{%
Facultad de Ciencias F\'isico-Matem\'aticas\\
Benem\'erita Universidad Aut\'onoma de Puebla, C.P. 72570, Puebla, Pue., Mexico.}%
\date{\today}

\begin{abstract}

Using the massive helicity formalism we calculate the five-body average square amplitude of the decays $\tau^{-}\to \ell_i^{-} \ell_i^{+} \ell_j^{-} \bar{\nu}_{j} \nu_{\tau}$ $(\ell=e,\,\mu)$ within the Standard Model (SM), we then introduce a dimension-five effective vertex $\Gamma^{\tau\tau \gamma}$  in order to determine the feasibility of imposing limits on the tau anomalous magnetic dipole moment ($a_\tau$) via the current or future experimental measurements of the branching ratio for the decay $\tau^{-}\to e^{-} e^{+} e^{-} \bar{\nu}_{e} \nu_{\tau}$. \\


\end{abstract}

\pacs{Valid PACS appear here}
\maketitle


\section{\label{sec:level1} Introduction}
We found that the SM five body decays $\tau^{-}\to \ell_i^{-} \ell_i^{+} \ell_j^{-} \bar{\nu}_{j} \nu_{\tau}$ whose branching ratios, at the tree level, have been calculated by Dicus and Vega \cite{Dicus}, Volobouev of the CLEO Collaboration \cite{CLEO} and  more recently by L\'opez Castro et. al. \cite{Roig}. However \cite{Dicus} and \cite{Roig} differ from the predictions of \cite{CLEO} whose various branching ratios are $\sim7\%$ higher.  The CLEO II experiment has searched for the $\tau^-\to e^- e^+ e^- \bar{{\nu}_{e}} \nu_{\tau}$ decay, whose measured value is \cite{PDG}:  
\begin{equation}
BR(\tau^-\to e^- e^+ e^- \bar{{\nu}_{e}} \nu_{\tau})=2.7^{+1.5+0.4+0.1}_{-1.1-0.4-0.3}\times 10^{-5}.
\end{equation}
The statistical, systematic and background uncertainties are $\sim70\%$ but forthcoming BELLE measurements are expected to improve to $\sim2.7\%$ and $\sim6.5\%$ on the statistical and systematic uncertainties \cite{junya}, respectively. The importance of our independent calculation is that it is useful to overcome this discrepancy. Furthermore, we analyze the possibility of limiting the $a_\tau$ value by using current and future experimental measurements by BELLE or other future collaborations.

Leptons offer some of the cleanest signals that can be obtained in collider physics experiments
and might also provide new  insight into physics beyond the standard model (SM). In particular, with the
yet unsolved origin of the discrepancy between the experimental and theoretical SM prediction of the muon anomalous magnetic dipole moment $a_\mu$. Although much work
has already been devoted to $a_\mu$, the $a_{\tau}$  has recently become the source of theoretical and experimental interest,
which is why in the present work we suggest a tau decay as a means to obtain further insight into $a_{\tau}$.
The current lower and upper bounds on $a_{\tau}$, $-0.052\leq a_{\tau}\leq 0.013$ with 95$\%$ C.L.
\cite{DELPHI}, were obtained via the process
$e^+ e^- \to e^+ e^- \tau^+ \tau^-$ by the DELPHI collaboration. These bounds differ from the theoretical value predicted by the SM by one order of magnitude:
$a_{\tau}^{\rm Theor.}=1177.21(5)\times 10^{-6}$ \cite{SMMomMagTAU}. Measurements of the electron and muon Anomalous 
Magnetic Dipole Moments (AMDM) were obtained by
means of spin precession experiments, however, in the case of the tau
lepton this class of measurements are troublesome due to its short lifetime, $(290.3\pm0.5)\times10^{-15} s$.
Even though there is a plethora of tau decays, only a few of them are viable candidates to
constrain $a_{\tau}$. Whereas two- and three-body decays do not involve the $\gamma\tau^-\tau^+$ coupling, the
$\tau^{\pm}\to\ell^{\pm}\bar{\nu_{\ell}}\nu_{\tau}\gamma$ decay, as suggested in Ref. \cite{tauTOfourbodies}, is
constrained by the tau lifetime and  is only sensitive to large values of $a_{\tau}$. The use of tau decays to constrain $a_\tau$ is particularly relevant because the Belle-II experiment is
expected to produce about $10^9$ tau leptons annually,  which greatly exceeds the previous
CLEO-II experiment, in which there were  $3\times 10^6$ produced tau leptons. 

\begin{figure}[h!]
\centering
{\includegraphics[width=15cm]{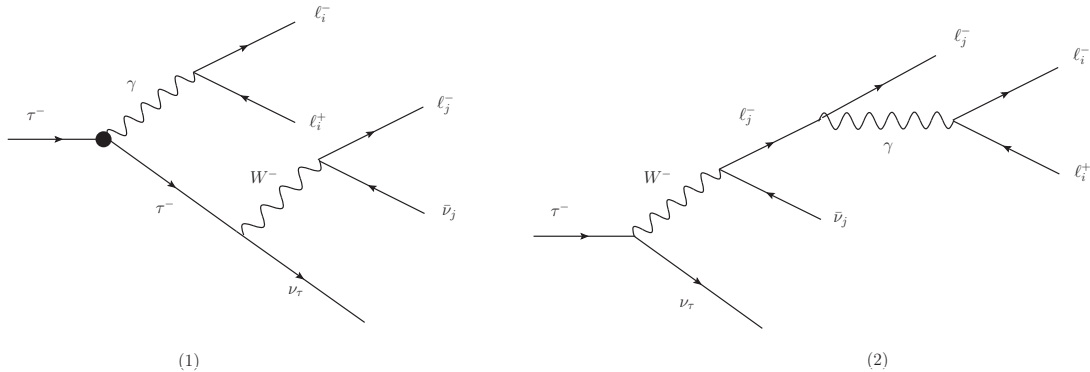}}
\caption{Dominant Feynman diagrams in the unitary gauge for the $\tau^{-}\to \ell_i^{-} \ell_i^{+} \ell_j^{-} \bar{\nu}_{j} \nu_{\tau}$ decay in the SM. The dot represents the QED contribution along with an extra contribution  from the tau AMDM. When $i\ne j$ there are two additional diagrams where the particles with the indices $i$ and $j$ are exchanged. Those diagrams will be denoted by the numbers (3) and (4) in our calculation below.\label{FeynDiag}}
\end{figure}

The dominant Feynman diagrams for this decay within the SM are shown in Fig. \ref{FeynDiag}, where the dot represents  the QED contribution along with an extra contribution  from the tau AMDM. All other tree-level diagrams give a negligible contribution.

In this work we will consider this branching ratio  to obtain a  bound on $a_\tau$. To this aim we will consider the following  effective 
vertex of the photon to a charged lepton pair respecting Lorentz invariance:
\begin{eqnarray}
\label{effectivelagrangian}
\Gamma^{\gamma\ell\ell}(q)&=&ie\,\bar\ell(p_2)\left[F_V^\gamma\gamma^\mu+\sigma_{\mu\nu}q^{\nu}\left(iF_{M}^{\gamma}
+F_{E}^{\gamma}\gamma_{5}\right)\right]\ell(p_1)A_\mu,
\end{eqnarray}
where $q=p_1-p_2$ is the photon transferred four-momentum.  Here ${F_V^\gamma}$ is the tau electric charge form factor (${F_V^\gamma}=1$ at the tree level), whereas the five-dimensional
$CP$-conserving and $CP$-violating  terms correspond to the static anomalous magnetic dipole moment $a_{\ell}$ and the electric dipole moment $d_{\ell}$, which are given by:
\begin{eqnarray}
a_{\ell}^{W} & = & -2m_{\ell}F_{M}^{\gamma}(q^2=0),\\
d_{\ell}     & = & -eF_{E}^{\gamma}(q^2=0).
\end{eqnarray}
We will assume  $CP$ invariance and take $F_E^{\gamma}$=0.

The rest of this presentation is organized as follows. In Section \ref{sec:amplitude} we will present   the unpolarized square amplitude for the  $\tau^{-}\to e^{-} e^{+} e^{-} \bar{\nu}_{e} \nu_{\tau}$ decay via the massive helicity formalism \cite{dittmaier}, which considerably simplifies the calculation. The numerical integration of the $\tau^{-}\to e^{-} e^{+} e^{-} \bar{\nu}_{e} \nu_{\tau}$ decay width and the resulting bound on the $a_{\tau}$ are presented in Sec. \ref{sec:numerical}, whereas Sec. \ref{sec:conclusion} is devoted to the conclusions and outlook. Details of the calculation are presented in Appendix \ref{sec:appendix}.


\section{\label{sec:amplitude} Unpolarized square amplitude}

We will calculate the average square amplitude of the $\tau^{-}\to \ell_i^{-} \ell_i^{+} \ell_j^{-} \bar{\nu}_{j} \nu_{\tau}$  decay using the massive helicity formalism for which usual treatments, such as \cite{dixon} or \cite{elvang}, only deal with the massless case. Here we need to take into account the mass of the tau lepton, so must go further. In particular, in \cite{dittmaier} the two main ways to deal  with massive helicity amplitudes are presented, although with a somewhat old-fashioned notation. We use here the approach that consists in performing a light-cone decomposition to write the four-momentum of a massive particle as a linear combination of two light-like momenta. For a detail account we refer the interested reader to Ref. \cite{proceedings},  where this formalism is presented in a self-contained manner (with opposite metric signature convention). We will present a brief outline here for convenience.

Let $p$ be the four-momentum of a massive particle of mass $m > 0$, which can always be written in terms of two light-like ones $q^\mu$ and $r^\mu$ as
\begin{equation}
	p^\mu = r^\mu + \frac{m^2}{2 p \cdot q} q^\mu ,
	\label{prq}
\end{equation}
where $q$ satisfies $p \cdot q \neq 0$ but is otherwise arbitrary, and $r$ is defined through \eqref{prq}. The positive frequency momentum space Dirac equation has two linearly independent solutions, which we label with the subindices $+$ and $-$. It is easily checked that they are given (maybe up to a phase) by
\begin{equation}
	u_+ = | r \ra + \frac{m}{[rq]} | q ] \quad , \quad u_- = | r ] + \frac{m}{\la rq \ra} | q \ra ;
\end{equation}
while for the negative frequency we have
\begin{equation}
	v_+ = | r ] + \frac{m}{\la qr \ra} | q \ra \quad , \quad v_- = | r \ra + \frac{m}{[qr]} | q ] .
\end{equation}
Here $| a]$ and $| b \ra$ are 2-component Weyl spinors linked to the light-like momenta $a_\mu$ and $b_\mu$ \cite{proceedings}.
We will use these solutions to calculate the helicity amplitudes $\mathcal{M}_a^{h_1 h_2 h_3 h_4 h_5 h_6}$ for the $\tau^{-}(p_1)\to \ell_i^{-}(p_2) \ell_i^{+}(p_3) \ell_j^{-}(p_4) \bar{\nu}_{j}(p_5) \nu_{\tau}(p_6)$ decay, where $h_k$ is the helicity of the particle $k$ with four-momentum $p_k$, and the subindex $a$ labels each Feynman diagram of Fig. \ref{FeynDiag}: it runs from 1 to 4 in the case that the charged leptons $\ell_i^{-}$ and $\ell_j^{-}$  are identical, such as in the $\tau^{-}\to e^{-} e^{+} e^{-} \bar{\nu}_{e} \nu_{\tau}$ decay, but it only runs from from 1 to 2 if  $\ell_i^{-}$ and $\ell_j^{-}$ are distinguishable.

To  determine the individual non-zero helicity amplitudes we will neglect the mass of the outgoing charged leptons, which is a good approximation as their mass is negligible as compared to the tau mass. In this limit, each amplitude will vanish unless the helicities of particles 4, 5 and 6 have certain fixed signs because of its spinor index structure. After doing that, we present the sum of the squared modulus of the helicity amplitudes and then all of the  interferences.


In the unitary gauge, the amplitude for the first Feynman diagram  of Fig. \ref{FeynDiag}, dubbed (1), is
\begin{eqnarray}
	\mathcal{M}_1 &=& G_1 \bar{u} (4) \gamma^\mu P_L v(5) \left( g_{\mu\nu} - \frac{{k_3}_\mu {k_3}_\nu }{m_W^2} \right) \bar{u} (6) \gamma^\nu P_L ( \slashed{k}_2 + m_\tau ) ( \gamma^\rho {F_V^\gamma} + i \sigma^{\rho\beta} {k_1}_\beta F_M^\gamma ) u (1)\nonumber\\&\times& \bar{u} (2) \gamma_\lambda v(3) ,
\end{eqnarray}
where we use the short-hand notation $(i)\equiv (p_i)$ and the four-momenta $k_i$ are the ones of the virtual particles of this diagram. Note that we are using the effective interaction  (\ref{effectivelagrangian}) for the $\gamma\bar\tau\tau$ vertex but in our calculation below we will use the tree-level value $F_V^\gamma=1$. In addition
\begin{equation}
	G_1 \equiv \frac{g_W^2 e^2}{8 p_{23} ( p_{23} - p_{12} - p_{13} ) ( 2 p_{45} - m_W^2 )},
\end{equation}
with  $p_{ij}=p_i\cdot p_j$.
The helicity structure of the amplitude fixes $h_4=-$, $h_5=+$, and $h_6=-$. We then write
\begin{equation}
	\mathcal{M}_1 = G_1 \la 4 | \gamma_\mu | 5 ] \la 6 | \gamma^\mu ( \slashed{k}_2 + m_\tau ) \gamma^\nu \left( {F_V^\gamma} - {F_M^\gamma} \slashed{k}_1 \right) u (1) \bar{u} (2) \gamma_\nu v (3).
\end{equation}
We now choose $q_1 \equiv p_3$ and define
\begin{gather}
	D_1 \equiv \frac{p_{13}}{p_{23}} {F_V^\gamma} - m_\tau {F_M^\gamma} , \\
	E \equiv m_\tau {F_V^\gamma} - 2 p_{13} {F_M^\gamma} ,
\end{gather}
thus the individual helicity amplitudes are given by
\begin{equation}
	\mathcal{M}_1^{++--+-} = 4 G_1 \frac{ \la 64 \ra [23] }{ [r3] } \left( E ( [54] \la 43 \ra + [56] \la 63 \ra ) - m_\tau D_1 \la 32 \ra [25] \right) ,
\end{equation}
\begin{equation}
	\mathcal{M}_1^{+-+-+-} = 4 G_1 \la 64 \ra \la r2 \ra \left( m_\tau {F_V^\gamma} [35] - {F_M^\gamma} [23] ( [54] \la 42 \ra + [56] \la 62 \ra ) \right) ,
\end{equation}
\begin{equation}
	\mathcal{M}_1^{-+--+-} = 4 G_1 \la 64 \ra [2r] \left( {F_V^\gamma} ( [54] \la 43 \ra + [56] \la 63 \ra ) + m_\tau {F_M^\gamma} \la 32 \ra [25] \right) ,
\end{equation}
and
\begin{equation}
	\mathcal{M}_1^{--+-+-} = 4 G_1 \frac{ \la 64 \ra }{ \la r3 \ra} \left( 2 p_{23} D_1 ( [54] \la 42 \ra + [56] \la 62 \ra ) - m_\tau E \la 23 \ra [35] \right) .
\end{equation}
To obtain the helicity amplitudes of the the third Feynman diagram, obtained from the first one after the exchange of identical particles,  we simply exchange the momenta and helicities of particles 2 and 4, both in the helicity amplitudes and in the definitions of $D_1$ and $G_1$, leading to new coefficients that we denote as $D_3$ and $G_3$, respectively.

By an analogous procedure, we obtain for the  Feynman diagram (2) of Fig. \ref{FeynDiag}
\begin{gather}
	\mathcal{M}_2^{++--+-} = 4 m_\tau H_2 \frac{ [53] \la 34 \ra }{[r3]} ( \la 63 \ra [32] + \la 64 \ra [42] ) , \\
	\mathcal{M}_2^{+-+-+-} = 4 m_\tau H_2 \frac{ \la 24 \ra [53] }{[r3]} ( \la 62 \ra [23] + \la 64 \ra [43] ) , \\
	\mathcal{M}_2^{-+--+-} = 4 H_2 \la 34 \ra [5r] ( \la 63 \ra [32] + \la 64 \ra [42] ) , \\
	\mathcal{M}_2^{--+-+-} = 4 H_2 \la 24 \ra [5r] ( \la 62 \ra [23] + \la 64 \ra [43] ) ;
\end{gather}
where
\begin{equation}
	H_2 \equiv \frac{ g_W^2 e^2 }{ 8 p_{23} \left( p_{23} + p_{24} + p_{34} \right) \left( m_\tau^2 - m_W^2 - p_{16} \right) } .
\end{equation}
We straightforwardly obtain the corresponding amplitudes for the fourth diagram after the exchange of the momenta and helicities of particles 2 and 4.

We now factor all the dependence on the form factors ${F_V^\gamma}$ and ${F_M^\gamma}$ by defining the following coefficients:
\begin{gather}
	A_1 \equiv 2 p_{12} {F_V^\gamma}^2 + \frac{p_{23}}{p_{13}} ( E^2 + m_\tau^2 {F_V^\gamma}^2 ) , \\
	B_1 \equiv 2 p_{12} {F_M^\gamma}^2 + \frac{p_{23}}{p_{13}} ( D_1^2 + m_\tau^2 {F_M^\gamma}^2 ) , \\
	C_1 \equiv 2 p_{12} {F_V^\gamma} {F_M^\gamma} + \frac{p_{23}}{p_{13}} ( m_\tau^2 {F_V^\gamma} F_M - D_1 E ) ,
\end{gather}
from which we obtain the sum of the squared helicity amplitudes of the first diagram:
\begin{align}
	\sum_h \left| \mathcal{M}_1 \right|^2 &= 128 G_1^2 p_{46}
	\Bigg[ A_1 \left( \frac{m_\tau^2}{2} p_{35} + p_{34} p_{45} + p_{36} p_{56} + p_{3456} \right)\nonumber\\
	& + 2 p_{23} B_1 \left( \frac{m_\tau^2}{2} p_{25} + p_{25} p_{45} + p_{26} p_{56} + p_{2654} \right)
	 + 2 m_\tau p_{23} C_1 ( p_{45} + p_{56} ) \Bigg],
\end{align}
where $p_{i_1i_2\cdots i_N}$ is defined in Appendix \ref{sec:appendix}. For the second diagram we get
\begin{align}
	\sum_h \left| \mathcal{M}_2 \right|^2 &= 256 H_2^2  \left( m_\tau^2 \frac{p_{35}}{p_{13}} + p_{15} \right)
	\left[ p_{24} ( p_{23} p_{26} + p_{34} p_{46} + p_{3264} )\right.\nonumber\\&\left.+ p_{34} ( p_{23} p_{36} + p_{24} p_{46} + p_{2364} )  \right] .
\end{align}
The corresponding expression for the fourth diagram is attained by exchanging $p_2$ and $p_4$.

On the other hand, the non-zero interferences are given by

\begin{equation}
	\mathcal{I}_{12} = 64 G_1 H_2 \left[ 2 {F_V^\gamma} \left( \mathcal{Q}^{(0)}_{12} + \frac{m_\tau^2}{p_{13}} \mathcal{Q}^{(2)}_{12} \right) - m_\tau {F_M^\gamma} \mathcal{R}^{(0)}_{12} \right] ,
\end{equation}

\begin{align}
	\mathcal{I}_{13} &= 64 G_1 G_3  \Bigg[ {F_V^\gamma}^2 \left( p_{13} \mathcal{Q}^{(0)}_{13} + m_\tau^2 \mathcal{Q}^{(2)}_{13} + \frac{m^4}{p_{13}} \mathcal{Q}^{(4)}_{13} \right) + 2 {F_M^\gamma}^2 \left( \mathcal{R}^{(0)}_{13} + m_\tau^2 \mathcal{R}^{(2)}_{13} \right) \nonumber\\
&- m_\tau {F_V^\gamma} {F_M^\gamma} \left( \mathcal{P}^{(0)}_{13} + m_\tau^2 \mathcal{P}^{(2)}_{13} \right) \Bigg] ,
\end{align}

\begin{equation}
	\mathcal{I}_{14} = 64 G_1 H_4 \left[ 2 {F_V^\gamma} \left( \mathcal{Q}^{(0)}_{14} + \frac{m_\tau^2}{p_{13}} \mathcal{Q}^{(2)}_{14} \right) + m_\tau {F_M^\gamma} \mathcal{R}^{(0)}_{14} \right] ,
\end{equation}
and
\begin{equation}
	\mathcal{I}_{24} = - 512 H_2 H_4 p_{15} p_{24} \left( p_{23} p_{26} + p_{34} p_{46} + p_{2346} \right) ;
\end{equation}
where $\mathcal{I}_{ij}$ stands for the interference between the amplitudes of diagrams $i$ and $j$.  Explicit expressions for  the $\left( \mathcal{P}, \mathcal{Q}, \mathcal{R} \right) ^{(p)}_{ij}$ functions are given in Appendix \ref{sec:appendix}.

The full unpolarized squared amplitude for the $\tau^{-}\to e^{-} e^{+} e^{-} \bar{\nu}_{e} \nu_{\tau}$ decay is given by
 \begin{equation}
|\overline{\mathcal{M}}|^2= \frac{1}{2} \left( \sum_{i=1}^4\sum_h \left| \mathcal{M}_i \right|^2+\mathcal{I}_{12}+\mathcal{I}_{13}+\mathcal{I}_{14}+\mathcal{I}_{24} \right) ,
 \end{equation}

It is worth noting that this method is straightforward and  yields compact results  easy to handle in the numerical integration.


\section{\label{sec:numerical} Numerical results}
We now turn to compute the branching ratio of the tau  five-body decay $\tau^{-}\to e^{-} e^{+} e^{-} \bar{\nu}_{e} \nu_{\tau}$. The decay width is given by the usual formula
\begin{equation}
\label{decaywidth}
 \Gamma(\tau^{-}\to e^{-} e^{+} e^{-} \bar{\nu}_{e} \nu_{\tau})=\frac{(2\pi)^4}{2m_\tau}
 \int \prod_{i=1}^5 \frac{d^3p_i}{16\pi^3 E_i}|\overline{\mathcal{M}}|^2\delta^4\left(p_1-\sum_{i=2}^6 p_i\right).
\end{equation}
After dividing by the tau total width $\Gamma_\tau=1/\tau_\tau$  we obtain the corresponding branching ratio
\begin{equation}
 BR(\tau^{-}\to e^{-} e^{+} e^{-} \bar{\nu}_{e} \nu_{\tau})=\frac{\Gamma(\tau^{-}\to e^{-} e^{+} e^{-} \bar{\nu}_{e} \nu_{\tau})}{\Gamma_\tau}.
 \end{equation}

According to Ref.\cite{EspFase} the four-momenta of the involved particles in a five-body decay can be related to eight independent Lorentz invariant parameters through the relations:
\begin{equation}\label{varcine1}
s_1=(p_1-p_4)^2,\, s_2=(p_1-p_4-p_2)^2,\, s_3=(p_1-p_4-p_2-p_3)^2,
\end{equation}
\begin{equation}
u_1=(p_1-p_2)^2,\,u_2=(p_1-p_3)^2,\,u_3=(p_1-p_5)^2,
\end{equation}
\begin{equation}\label{varcine2}
t_2=(p_1-p_2-p_3)^2,\,t_3=(p_1-p_2-p_3-p_5)^2.
\end{equation}
The phase-space integral  (\ref{decaywidth}) was numerically computed over these kinematic variables via Monte-Carlo integration by using the VEGAS routines \cite{Lepage}. A cross-check of our results was done by implementing the electromagnetic vertex in the CalcHEP package \cite{Belyaev:2012qa}, which performs all the numerical calculation.

We first consider that the impact of the magnetic form factor is negligible, i.e. we use $F_M^{\gamma}=0$ and  make a comparison of our numerical results for the  widths of the allowed $\tau^-\to \ell_i^- \ell_j^+ \ell_i^- \bar{{\nu}_{\ell}}_j \nu_{\tau}$ decays with those obtained in previous studies. The results are shown in Table \ref{Results}.
The uncertainties arise from the numerical integration. Our results are in good agreement with these predictions, though are closer to those of Ref. \cite{Roig}, which could be attributed to the fact that we  used the same values of the tau mass and mean lifetime. Additionally, for completeness, we computed the $\mu^-\to e^- e^+ e^- \bar{\nu_e} \nu_{\mu}$ for which we obtain $BR(\mu^-\to e^- e^+ e^- \bar{\nu_e} \nu_{\mu})=(3.599\pm0.002)\times10^{-5}$.

\begin{table}[h!]
\centering
\def\arraystretch{1.5}
\begin{tabular}{|c|c|c|c|c|}
\hline
{{}Branching ratio} & {{}Ref{}\cite{Dicus}{}} & {{}Ref. \cite{CLEO}{}} & {{}Ref. \cite{Roig}{}} & {\bf{}Our results}\tabularnewline
\hline
\hline
{{}$\frac{BR(\tau^-\to e^-e^+e^-\bar{\nu_{e}}\nu_{\tau})}{10^{-5}}$} & {{}$4.15\pm0.06$} & {{}$4.457\pm0.006$} & {{}$4.21\pm0.01$} & {{}$4.22\pm0.02$}\tabularnewline
\hline
{{}$\frac{BR(\tau^-\to e^-\mu^+\mu^-\bar{\nu_{e}}\nu_{\tau})}{10^{-7}}$} & {{}$1.257\pm0.003$} & {{}$1.347\pm0.002$} & {{}$1.247\pm0.001$} & {{}$1.246\pm0.002$}\tabularnewline
\hline
{{}$\frac{BR(\tau^-\to\mu^-e^+e^-\bar{\nu_{\mu}}\nu_{\tau})}{10^{-5}}$} & {{}$1.97\pm0.02$} & {{}$2.089\pm0.003$} & {{}$1.984\pm0.004$} & {{}$1.987\pm0.003$}\tabularnewline
\hline
{{}$\frac{BR(\tau^-\to\mu^-\mu^+\mu^-\bar{\nu_{\mu}}\nu_{\tau})}{10^{-7}}$} & {{}$1.190\pm0.002$} & {{}$1.276\pm0.005$} & {{}$1.183\pm0.001$} & {{}$1.184\pm0.001                                                                                                                                                                                               $}\tabularnewline
\hline
\end{tabular}
\caption{Branching ratios for the allowed $\tau^-\to \ell_i^- \ell_j^+ \ell_i^- \bar{{\nu}_{\ell}}_j \nu_{\tau}$ decays in the SM. } \label{Results}
\end{table}

\subsection{Effect of $F_M^{\gamma}$ on the  $\tau^-\to e^-e^+e^-\bar{\nu_{e}}\nu_{\tau}$ decay}
\begin{figure}[h]
\begin{centering}
{\includegraphics[width=8cm]{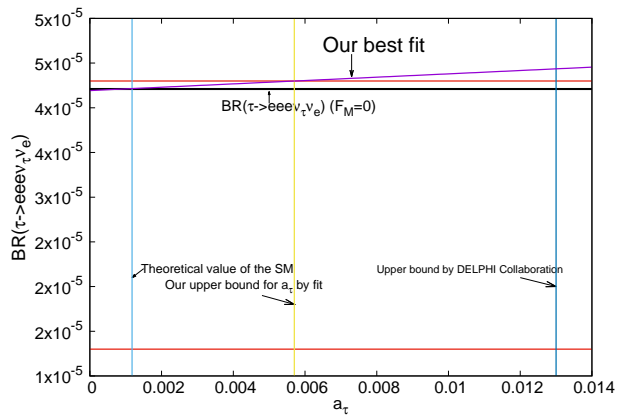}}
\caption{$BR(\tau^-\to e^- e^+ e^- \bar{{\nu}_{e}} \nu_{\tau})$ as a function of $a_{\tau}$ (purple line). The horizontal red lines are the 95\% C.L. limits on the experimental measurement of the five-body decay, of which the central value is $2.7\times10^{-5}$. Whereas the horizontal black line is the tree-level SM prediction ($F_M=0$). On the other hand, the vertical lines are the SM prediction of $a_\tau$(blue line), our bound (yellow line), and the upper DELPHI bound (dark blue line).   \label{branching}}
\par\end{centering}
\end{figure}

We will now analyze the impact of the dipole term $F_M^{\gamma}$  on the branching ratio of the tau five-body decay.
The behavior of  $BR(\tau^-\to e^-e^+e^-\bar{\nu_{e}}\nu_{\tau})$  as a function of $a_{\tau}$ is shown in Fig. \ref{branching}. The horizontal red lines corresponds to the 95\% C.L. interval obtained from the  experimental measurement of $BR(\tau^-\to e^- e^+ e^- \bar{{\nu}_{e}} \nu_{\tau})$ \cite{PDG}, and the  horizontal black line  corresponds to our calculation for the tree-level SM prediction, i.e. $BR(\tau^-\to e^- e^+ e^- \bar{{\nu}_{e}} \nu_{\tau})=(4.22\pm 0.02)\times 10^{-5}$. The purple curve corresponds to our prediction for the branching ratio as a function of $a_{\tau}$. Assuming that there are no extra  contribution rather than that due to $a_\tau$ we can conclude that the points of the purple curve falling above the upper experimental  bound on $BR(\tau^-\to e^- e^+ e^- \bar{{\nu}_{e}} \nu_{\tau})$ would be excluded, thereby yielding the  bound $a_{\tau}\le 0.0056$. This allow us to gain an improvement over the current upper bound by DELPHI: $a_{\tau}^{\rm DELPHI}\leq0.013$.  Since there is a slight dependence of $BR(\tau^-\to e^- e^+ e^- \bar{{\nu}_{e}} \nu_{\tau})$  on $a_\tau$,  this method is not useful to set a lower bound on it with current measurements. 

\begin{figure}[h]
\begin{centering}
{\includegraphics[width=6cm, angle=270]{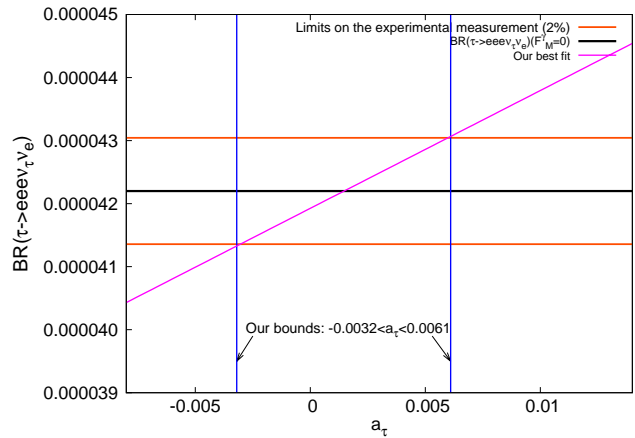}}
\caption{With an improvement of the uncertainties at the 2\% level in
$BR(\tau^-\to e^- e^+ e^- \bar{{\nu}_{e}} \nu_{\tau})$, our best fit prediction of $a_\tau$ (purple line), along with the improved measurements (red lines) allow us to ascertain that $-0.0032\leq a_{\tau}\leq 0.0061$.  The horizontal black line is the tree-level SM prediction ($F_M=0$).  \label{branching2}}
\par\end{centering}
\end{figure}

Forthcoming measurements might reach a statistical uncertainty of $2.7\%$ and a systematic uncertainty of $6.5\%$. Which we found is not good enough so that we may obtain  improved bounds, since our lower bound would only improve to a level of $\sim$-0.02. But new runs at BELLE II or future experiments we hope might offer more precise measurements, which is why we optimistically assume that if the statistical and systematic uncertainty is improved to the 2\% level, we expect to find that the bound of $a_\tau$ would fall in the range of $-0.0032\leq a_{\tau}\leq 0.0061$, see Fig. \ref{branching2}. 

Other approaches similarly yield stringent bounds on $a_\tau$ through electroweak precision data (EWPD) and the experimental data for the $e^+ e^- \to \tau^+ \tau^-$:  $-0.004\leq  a_{\tau}\leq 0.006$ \cite{Escribano}, $-0.007\leq a_{\tau}\leq 0.005$ \cite{Springberg} and more recently $-0.007\leq a_{\tau}\leq 0.004$ \cite{Epifanov}. While on the theoretical side, other model extensions predict values for $a_\tau$ of the order of $\mathcal{O}(10^{-9}-10^{-6})$ \cite{MoyTav}-\cite{MSSM}, which could be useful in the case of a discrepancy between the experimental measurement and the SM prediction.


\section{\label{sec:conclusion} Conclusions}

In this work we present the results of a numerical calculation for the predictions of the branching ratio of the allowed $\tau^{-}\to \ell_i^{-} \ell_i^{+} \ell_j^{-} \bar{\nu}_{j} \nu_{\tau}$ decays, which are consistent with previous results reported by Dicus and Vega and L\'opez Castro et. al. In addition, we use the current experimental measurement on the branching ratio of the $\tau^-\to e^- e^+ e^- \bar{{\nu}_{e}} \nu_{\tau}$ decay to obtain an upper limit on the tau anomalous magnetic dipole moment by using an effective electromagnetic vertex including a dipole term. We find that the effect  of the magnetic dipole form factor $F_M^{\gamma}$ on the reported value by CLEO II colaboration for the branching ratio of the $\tau^-\to e^- e^+ e^- \bar{{\nu}_{e}} \nu_{\tau}$ decay allows to extract an upper bound  of $a_{\tau}\le 0.0056$, which is below the current upper bound by the DELPHI collaboration. With the next measurements by BELLE II results are expected with $\sim2.7\%$ and $\sim6.5\%$ of statistical and systematic uncertainties. Under this scenario, we find that the effect of $F_M^{\gamma}$ on the $BR(\tau^-\to e^+e^+e^-\bar{\nu}_e \nu_{\tau})$ is such that  the lower bound is given by $-0.02\leq a_{\tau}$ but it is not possible to extract a good upper bound. Finally, assuming an accurate measurement of the $2\%$ of statistical and systematic uncertainties, it is found that our best bounds are $-0.0032\leq a_{\tau}\leq 0.0061$.

\begin{acknowledgments}
We acknowledge support from CONACYT (M\'exico).
\end{acknowledgments}


\appendix
\section{\label{sec:appendix} Conventions for the massive helicity formalism and interference terms for the $\tau^-\to e^- e^+ e^- \bar{{\nu}_{e}} \nu_{\tau}$ unpolarized square amplitude.}


We use the abbreviated notation
\begin{equation}
	p_{i_1 i_2 \dots i_N} \equiv \frac{1}{2} \text{Re} \left( [ i_1 i_2 ] \la i_2 i_3 \ra \dots [ i_{N-1} i_N ] \la i_N i_1 \ra \right)
	\label{p_def}
\end{equation}
(where $\text{Re} (z)$ stands for the real part of the complex number $z$), with $N$ even. It is easy to show that
\begin{equation}
	p_{i_1 i_2 \dots i_N} = \frac{1}{4} {i_1}_{\mu_1} {i_2}_{\mu_2} \dots {i_N}_{\mu_N} \text{Tr} \left( \gamma^{\mu_1} \gamma^{\mu_2} \dots \gamma^{\mu_N} \right) .
\end{equation}
It is well known that, for $N$ even, the trace of the product of $N$ gamma matrices is proportional (the proportionality constant being $4$) to $(N-1)!!$ terms, each of which is the product of $N/2$ metric tensors. In the particular case of only two indices we write $p_{ab}$ as an abbreviation of $p_a \cdot p_b$ even when the momenta $p_a$ and $p_b$ are not null. When there are more indices the individual momenta must be light-like. \newline
\indent Several properties can be obtained immediately. For instance, cyclicity of the trace translates into cyclicity of indices:
\begin{equation}
	p_{i_1 i_2 \dots i_{N-1} i_N} = p_{i_N i_1 i_2 \dots i_{N-1}} .
\end{equation}
Using the definition \eqref{p_def} we see that a multi-index $p$ with two equal adjacent indices vanishes because
\begin{equation}
	[pp] = \la pp \ra = 0 .
	\label{pp0}
\end{equation}
On the other hand, using $\lbrace \gamma^\mu , \gamma^\nu \rbrace = 2 g^{\mu\nu}$ we obtain the following general ``index-commuting formula'':
\begin{equation}
	p_{i_1 i_2 \dots i_{k-1} i_k i_{k+1} i_{k+2} \dots i_N} = 2 p_{i_k i_{k+1}} p_{i_1 i_2 \dots i_{k-1} i_{k+2} \dots i_N} - p_{i_1 i_2 \dots i_{k-1} i_{k+1} i_k i_{k+2} \dots i_N} .
	\label{commute_indices}
\end{equation}
\indent In a scattering (or decay) process with $L$ external legs, we have $L$ distinct four-momenta. If we calculate a squared amplitude and get a $p$ with more than $L$ indices, we use the \eqref{pp0} and \eqref{commute_indices} properties to write everything in terms of only $p$'s with at most $L$ indices, i.e. $L$ is an upper bound for the number of indices in an $L$ particle tree level process. In our case there are six external legs, therefore $p$'s with more than six indices shall not appear. \newline
\indent For the interferences we use the following expressions:
\begin{small}
\begin{align}
	\mathcal{Q}^{(0)}_{12} & = p_{24} \left( 2 p_{46} p_{3r54} + p_{2645r3} \right) + p_{34} \left( 2 p_{46} p_{2r54} + p_{2r5463} \right) \\
	& + p_{46} \left( p_{2r5634} + p_{243r56} \right) - p_{26} p_{2465r3} - p_{36} p_{2r5643} ,
\end{align}
\begin{align}
	\mathcal{Q}^{(2)}_{12} = & 2 p_{35} p_{46} p_{243r} + p_{23} \left( p_{34} p_{3645} - p_{36} p_{3564} \right) + p_{46} \left( p_{34} p_{2354} - p_{36} p_{2534} + p_{35} p_{2634} \right) \\
	+ & p_{13} \left( p_{26} p_{2534} + p_{34} p_{2563} + p_{46} p_{2534} - p_{24} p_{2536} - p_{36} p_{2543} - p_{46} p_{2453} \right) ,
\end{align}
\begin{align}
	\mathcal{R}^{(0)}_{12} = 4 & \Big[ p_{23} ( p_{34} p_{3645} - p_{36} p_{3564} )+ p_{46} ( p_{34} p_{2354} - p_{36} p_{2534} p_{35} p_{2634} ) \\
	& + p_{35} ( p_{46} p_{24r3} - p_{26} p_{2r34} + p_{24} p_{2r36} ) - p_{13} ( p_{24} p_{2536} - p_{26} p_{2534} + p_{46} p_{2453} ) \\
	& + p_{12} ( p_{46} p_{2435} - p_{34} p_{2635} + p_{36} p_{2435} ) - p_{25} ( p_{46} p_{243r} - p_{34} p_{263r} + p_{36} p_{243r} ) \\
	& - \frac{p_{24} p_{46}}{p_{13}} \left( p_{13} p_{2653} - p_{34} p_{253r} + p_{35} p_{243r} - 2 p_{23} p_{3r54} \right) \\
	& + p_{23} \left( p_{24} p_{2645} - p_{26} p_{2465} + p_{46} p_{2456} \right) \Big] ;
\end{align}
\begin{equation}
	\mathcal{Q}^{(0)}_{13} = 4 \left( 2 p_{26} p_{46} p_{56} + p_{46} p_{2654} - p_{24} p_{2645} + p_{26} p_{2465} \right) ,
\end{equation}
\begin{align}
	\mathcal{Q}^{(2)}_{13} = 2 & \left( p_{35} p_{264r} + p_{45} p_{2634} - p_{34} p_{2654} - p_{26} p_{3465} + p_{24} p_{6253} - p_{46} p_{2653} - p_{26} p_{4253} \right) ,
\end{align}
\begin{equation}
	\mathcal{Q}^{(4)}_{13} = p_{35} p_{2643} ,
\end{equation}
\begin{align}
	\mathcal{R}^{(0)}_{13} & = 4 p_{26} \left( p_{46} p_{56} p_{234r} + p_{23} p_{46} p_{254r} - p_{23} p_{45} p_{264r} \right) - 4 p_{24} \left( p_{23} p_{45} p_{264r} - p_{25} p_{46} p_{324r} + p_{23} p_{46} p_{524r} \right) \\
	& + 4 p_{46} \left( p_{12} p_{34} p_{2456} - \frac{1}{2} p_{24} p_{2r3456} \right) + 2 p_{24} p_{23564r} \left( p_{24} + p_{26} \right) ,
\end{align}
\begin{align}
	\mathcal{R}^{(2)}_{13} & = \frac{2 p_{23} p_{34}}{p_{13}} \left( 2 p_{26} p_{46} p_{56} + p_{46} p_{2654} + p_{26} p_{2465} - p_{24} p_{2645} \right) + 2 p_{13} p_{35} p_{2643} \\
	& + 2 p_{23} ( p_{45} p_{2634} - p_{34} p_{2654} - p_{26} p_{3465} ) - 2 p_{34} ( p_{26} p_{4253} - p_{24} p_{6253} + p_{46} p_{2653} ) ,
\end{align}
\begin{align}
	\mathcal{P}^{(0)}_{13} & = 4 ( p_{23} + p_{34} ) ( 2 p_{26} p_{46} p_{56} + p_{46} p_{2654} - p_{24} p_{2645} + p_{26} p_{2465} ) \\
	& + 2 p_{13} \left( p_{45} p_{2634} - p_{34} p_{2654} - p_{26} p_{3465} \right) - 2 p_{13} \left( p_{26} p_{4253} - p_{24} p_{6253} + p_{46} p_{2653} \right) \\
	& - 2 p_{24} \left( p_{264r35} + p_{23564r} \right) - 2 p_{26} p_{23564r} - 2 p_{46} p_{26534r} \\
	& - 4 p_{14} \left( p_{25} p_{2643} - p_{23} p_{2645} \right) - 4 p_{46} \left( p_{25} p_{24r3} - p_{23} p_{24r5} \right) ,
\end{align}
\begin{align}
	\mathcal{P}^{(2)}_{13} & = 4 p_{35} p_{2643} + 2 \frac{p_{34}}{p_{13}} \left( p_{23} p_{2645} - p_{25} p_{2643} \right) \\
	& + \frac{p_{23}}{p_{13}} ( p_{45} p_{2634} - p_{34} p_{2654} - p_{26} p_{3465} ) - \frac{p_{34}}{p_{13}} ( p_{26} p_{4253} - p_{24} p_{6253} + p_{46} p_{2653} ) ,
\end{align}
\begin{equation}
	\mathcal{Q}^{(0)}_{14} = p _{26} p_{2465r3} - p_{24} p_{2645r3} - p_{46} p_{243r56} - 2 p_{24} p_{46} p_{3r54} ,
\end{equation}
\begin{align}
	\mathcal{Q}^{(2)}_{14} & = p_{35} \left[ p_{24} p_{2r36} - ( p_{26} + 3 p_{46} ) p_{243r} \right] - p_{13} \left[ p_{24} p_{2536} - ( p_{26} + p_{46} ) p_{2453} - 2 p_{24} p_{35} p_{46} \right] ,
\end{align}
\begin{align}
	\mathcal{R}^{(0)}_{14} = & 4 \left[ p_{46} \left( p_{35} p_{24r3} - p_{13} p_{2453} \right) + p_{13} \left( p_{24} p_{2536} - p_{26} p_{2534} \right) + p_{35} \left( p_{24} p_{2r36} - p_{26} p_{2r34} \right) \right] \\
	+ & \frac{2p_{23}}{p_{13}} \left( p_{24} p_{26453r} - p_{26} p_{24653r} + p_{46} p_{2653r4} \right)	+ 4 p_{23} \left( p_{24} p_{2645} - p_{26} p_{2465} + p_{46} p_{2456} \right) \\
	+ & \frac{8 p_{24} p_{46}}{p_{13}} \left( p_{34} p_{253r} - p_{35} p_{243r} - p_{13} p_{2453} + \frac{1}{2} p_{23} p_{3r54} \right) ;
\end{align}
\end{small}
where the $p_{i_1 i_2 \dots i_{N-1} i_N}$ can be written in terms of the $p_{ab}$ by means of the following recursion formulas
\begin{align}
p_{abcd} &= p_{ab}p_{cd} - p_{ac}p_{bd} + p_{ad}p_{bc},\\
p_{abcdef} &= p_{ab}p_{cdef} - p_{ac}p_{bdef} + p_{ad}p_{bcef} - p_{ae}p_{bcdf} + p_{af} p_{bcde} .
\end{align}
Furthermore, the subindex $r$ is determined by 
\begin{align}
p_{3r} &= p_{13},\\
p_{ir} &= p_{1i}- \frac{m_\tau^2}{ 2p_{13}}p_{3i}, \qquad{\rm for}\quad  i = 2,4,5,6,
\end{align}


\end{document}